\documentclass{aa}
\usepackage{graphicx}
\usepackage{txfonts}
\begin{document}
   \title{Gaussian decomposition of \ion{H}{i} surveys}
   \subtitle{V. Search for very cold clouds}
   \author{U. Haud}
   \institute{Tartu Observatory, 61\,602 T\~oravere, Tartumaa, Estonia\\
              \email{urmas@aai.ee} }
   \date{Received \today; accepted \today}
   \abstract
      {In the previous papers of this series, we have decomposed into
         Gaussian components all the \ion{H}{i} 21-cm line profiles of
         the Leiden-Argentina-Bonn (LAB) database, and studied
         statistical distributions of the obtained Gaussians.}
      {Now we are interested in separation from the general database of
         the components the ``clouds'' of closely spaced similar
         Gaussians. In this paper we examine the most complicated case
         for our new cloud-finding algorithm -- the clouds of very
         narrow Gaussians.}
      {To separate the clouds of similar Gaussians, we start with the
         single-link hierarchical clustering procedure in
         five-dimensional (longitude, latitude, velocity, Gaussian width
         and height) space, but make some modifications to accommodate
         it to the large number of components. We also use the
         requirement that each cloud may be represented at any observed
         sky position by only one Gaussian and take into account the
         similarity of global properties of the merging clouds.}
      {We demonstrate that the proposed algorithm enables us to find the
         features in gas distribution, which are described by similar
         Gaussians. As a test, we apply the algorithm for finding the
         clouds of the narrowest \ion{H}{i} 21-cm line components.
         Using the full sky search for cold clouds, we easily detect the
         coldest known \ion{H}{i} clouds and demonstrate that actually
         they are a part of a long narrow ribbon of cold clouds. We
         model these clouds as a part of a planar gas ring, deduce their
         spatial placement, and discuss their relation to supernova
         shells in the solar neighborhood. Many other narrow lined
         \ion{H}{i} structures are also found.}
      {We conclude that the proposed algorithm satisfactorily solves the
         posed task. In testing the algorithm, we found a long ribbon of
         very cold \ion{H}{i} clouds and demonstrated that all the
         observed properties of this band of clouds are very well
         described by the planar ring model. We also guess that the
         study of the narrowest \ion{H}{i} 21-cm line components may be
         a useful tool for finding the structure of neutral gas in solar
         neighborhood.}

      \keywords{ISM: atoms~-- ISM: clouds~-- Radio lines: ISM}

   \maketitle

   \section{Introduction}

      In earlier papers of this series, we described the Gaussian
      decomposition of large \ion{H}{i} 21-cm line surveys (Haud
      \cite{Hau00}, hereafter Paper~I) and the usage of the obtained
      Gaussians for the detection of different observational and
      reductional problems (Haud \& Kalberla \cite{Hau06}, hereafter
      Paper~II), for the separation of thermal phases in the
      interstellar medium (ISM; Haud \& Kalberla \cite{Hau07}, hereafter
      Paper~III) and for the studies of intermediate and high velocity
      hydrogen clouds (IVCs and HVCs; Haud \cite{Hau08}, hereafter
      Paper~IV). A detailed justification for the use of Gaussian
      decomposition in such studies was provided in Paper III.
      Observational data for the decomposition were from the LAB
      database of \ion{H}{i} 21 cm line profiles, which combines the new
      revision (LDS2, Kalberla et al. \cite{Kal05}) of the
      Leiden/Dwingeloo Survey (LDS, Hartmann \cite{Har94}) and a similar
      Southern sky survey (IARS, Bajaja et al. \cite{Baj05}) completed
      at the Instituto Argentino de Radioastronomia. The LAB database is
      described in detail by Kalberla et al. (\cite{Kal05}). Our method
      of Gaussian decomposition generated 1\,064\,808 Gaussians for
      138\,830 profiles from LDS2 and 444\,573 Gaussians for 50\,980
      profiles from IARS.

      In Papers II-IV, we used every obtained Gaussian as a single
      entity, which is independent of all other Gaussians, and analyzed
      statistical distributions of their parameters. The obtained
      results indicated that different structures in the ISM could be
      recognized as density enhancements in the distribution of Gaussian
      parameters in the five-dimensional parameter space, or that the
      well defined Gaussians with similar parameters at least
      statistically define in the real space the related objects, which
      share the same physical state. The situation may be more
      complicated in the cases of heavily blended Gaussians in emission
      lines near the Galactic Plane. These earlier papers demonstrated
      also the importance of the Gaussian widths, the knowledge of which
      helps us separate the components, corresponding to different
      physical structures of the ISM or to the artifacts of the
      observations, reduction and the Gaussian decomposition itself. The
      third point, clear from the earlier studies, is that in reality
      many \ion{H}{i} structures, observable in sky, extend to much
      larger areas than that covered by a single beam of the radio
      telescope. This means that some Gaussians of the neighboring
      profiles may represent the features of a similar origin, they are
      not independent of the others, and may be grouped together to
      represent larger structures.

      In the present paper, we start studying these similarities and
      relations between the Gaussians -- we define clouds of similar
      Gaussians, which may (but need not) describe the real gas
      concentrations in the real space. In doing so, we must keep in
      mind that there are no precise definitions of the terms such as
      ``cloud'', ``clump'', or ``core'' (Larson \cite{Lar03}), and the
      physical reality of the clumps, found by different authors, has
      been a matter of controversial debates since the presentation of
      the first systematic attempts to identify any kind of gas clumps.
      Nevertheless, many papers are devoted to the study of clouds,
      clumps and cores in the ISM.

      As the structure in molecular clouds determines, in part, the
      locations, numbers, and masses of newly formed stars, specifically
      great effort has been invested in characterizing the structure of
      this gas. Such statistical analysis of the molecular line data has
      usually followed one of two paths (Rosolowsky et al.
      \cite{Ros08}). Either authors construct statistical descriptions
      of the emission from an entire molecular line data set, or segment
      the data into what they believe to be physically relevant
      structures, and study the distribution of properties in the
      resulting population of objects.

      Common examples of statistical analysis include fractal analysis
      (Elmegreen \& Falgarone \cite{Elm96}; Stutzki et al.
      \cite{Stu98}), $\Delta$-variance (Stutzki et al. \cite{Stu98};
      Bensch et al. \cite{Ben01}), correlation functions (Houlahan \&
      Scalo \cite{Hou90}; Rosolowsky et al. \cite{Ros99}; Lazarian \&
      Pogosyan \cite{Laz00}) and principle component analysis (Heyer \&
      Brunt \cite{Hey04}). Statistical analysis produces many
      interesting comparisons between and among data, but physical
      interpretation of the statistics can be complicated. The
      segmentation and identification techniques are favored in the
      cases where the emission is thought to be comprised of physically
      important substructures (Rosolowsky et al. \cite{Ros08}).

      The clumpy substructure of molecular clouds was first identified
      by eye (Blitz \& Stark \cite{Bli86}; Carr \cite{Car87}; Loren
      \cite{Lor89}; Nozawa et al. \cite{Noz91}; Lada et al.
      \cite{Lad91}; Blitz \cite{Bli93}; Dobashi et al. \cite{Dob96}).
      However, as a power law mass spectrum predicts an increasing
      number of smaller and smaller clumps, confusion is usually the
      limiting factor in clump identification by eye. It is thus highly
      desirable to use automated clump finding algorithms in the
      analysis of observed data, as the use of an algorithm allows to
      analyze the structure in a consistent and stable way (Kramer et
      al. \cite{Kra98}).

      The two applications of the segmentation approach that have most
      shaped molecular line astronomy are the clump identification
      algorithms GAUSSCLUMP by Stutzki \& G\"{u}sten (\cite{Stu90}) and
      CLUMPFIND by Williams et al. (\cite{Wil94}). GAUSSCLUMPS uses a
      least square fitting procedure to decompose the emission
      iteratively into one or more Gaussian clumps. CLUMPFIND associates
      each local emission peak and the neighboring pixels with one clump
      (similar to the usual eye inspection procedure). Although the
      basic concept of both algorithms is quite different, they give
      consistent results for the larger clumps, when used on the same
      data (Williams et al. \cite{Wil94}). In the lower mass range, the
      emission is assigned to additional smaller clumps in the Gaussian
      decomposition algorithm, and to the irregular extensions of the
      more massive clumps by CLUMPFIND. In both cases, an implicit
      assumption is made that the radial velocity coordinate can be
      replaced by the radial distance, but this assumption is not
      necessarily always satisfied (Ostriker et al. \cite{Ost01}).

      With \ion{H}{i} data Thilker (\cite{Thi98}) has used an algorithm,
      somewhat similar to GAUSSCLUMPS, which creates a predefined set of
      various possible clouds, and applied it to look for \ion{H}{i}
      bubbles blown by supernovas in external galaxies. The method,
      similar to CLUMPFIND, was used by de Heij et al. \cite{deH02} to
      automatically search for compact high velocity clouds (HVCs) in
      the Leiden/Dwingeloo Survey. However, instead of scanning for
      clouds along contours of constant intensity at varying levels,
      they used the gradient of the intensity field to determine the
      structure to which the pixels should be assigned.

      Later Nidever et al. (\cite{Nid08}) have used the Gaussian
      decomposition of the LAB profiles with the algorithm created
      according to the description of our decomposition program in Paper
      I. The obtained Gaussians are then used to disentangle overlapping
      \ion{H}{i} structures. They stress that by using Gaussians, it is
      possible to distinguish different \ion{H}{i} filaments even when
      they are overlapping in velocity. They state that in these
      situations the Gaussians trace structures that are real, and they
      may even hold physical information about the structures. They were
      successful in tracking tenuous structures through rather
      complicated environments even though the decomposition of those
      environments likely holds no physical meaning. The results were
      used to study the origin of the Magellanic Stream and its leading
      arm.

      In this paper, we would like to move a step further and present an
      automatic computer program for finding different continuous
      \ion{H}{i} features in the full decomposition of the LAB database.
      In the next section, we describe our algorithm for finding
      coherent structures in the large database of Gaussians, and in the
      following section, we apply this algorithm to look for the coldest
      \ion{H}{i} clouds in the Galaxy, which could be identified using
      the LAB data. This search has been used for testing the new
      program as we may expect the clouds of very narrow Gaussians to be
      the hardest to find for our algorithm. However, when doing the
      search, we must keep in mind that the widths of such narrow
      Gaussians most likely are not correct representations of the
      actual widths of the underlying \ion{H}{i} 21-cm emission lines.
      Due to both the finite optical depth of the lines and the velocity
      resolution of the LAB survey, the Gaussian widths used in this
      paper are only the upper limits for the actual line widths.
      Therefore, they cannot be used for the study of physical
      properties in the found clouds, but as the actual lines are even
      narrower than the corresponding Gaussians, we may still state that
      we are looking for very narrow lines and very cold gas clouds.

   \section{The cloud finding algorithm}

      \subsection{The problems}

      The task of finding the clouds of similar objects belongs to
      cluster analysis. As in previous papers of this series we have
      studied some distributions of Gaussians, considering all
      components more or less independent of each other, it is now
      natural to follow some agglomerative (bottom-up) hierarchical
      clustering procedure. If we are looking for clouds, whose
      parameters vary smoothly from one point to another, the most
      appropriate algorithm seems to be single-link clustering: we add
      to the existing cluster a new element, which is the closest to at
      least one of the elements of this cluster.

      To find the closest pairs, we need to know the distances between
      all Gaussians in 5-dimensional space (two sky coordinates and
      three Gaussian parameters for every component). However, here we
      face three problems. First of all, what is the distance between
      two Gaussians? After answering this, we may compute the distances,
      but need to store the results somewhere for the future use. As
      mentioned in the Introduction, we have 1\,509\,381 Gaussians and
      consequently 1\,139\,114\,746\,890 distances between them. If we
      want to store all these and for every distance also the
      identifications of the Gaussians, between which it is calculated,
      we need a storage of about 12.4~TBytes. Today this is clearly
      possible, but by no means reasonable.

      When the first two problems are more or less mathematical, the
      third one is physical. In the Galaxy, \ion{H}{i} has a rather
      complicated spatial and kinematic structure, and therefore the
      Gaussian decompositions of the profiles, particularly near the
      galactic plane, may be rather complicated and contain many
      different Gaussians per one profile. Running the preliminary
      versions of our cluster identification program demonstrated that
      without applying special restrictions, one dominating cloud
      started to emerge around the galactic plane from the first steps
      of the merging process and finally only these \ion{H}{i}
      structures were distinguishable, whose properties very strongly
      differ from those of general ISM. We would like to achieve rather
      opposite results: to separate all the pieces of more or less
      coherent structures and follow them as close as possible to other
      structures, but not to merge probably different features.
      Therefore, the simplest version of the single-link clustering is
      not acceptable, and we have to modify the clustering process
      somehow to better suit our interests.

      \subsection{Similarity of Gaussians}

      An important step in any cluster finding process is the selection
      of a distance measure, which will determine how the similarity of
      two elements is calculated. In our case, this problem may be
      divided into two subproblems: one is the distance of the observed
      profiles on the sky and another the comparison of the shapes of
      the Gaussians. The first subproblem has a standard solution --
      according to spherical trigonometry, the distance, $e$, between
      two observing directions with galactic coordinates $(l_1, b_1)$
      and $(l_2, b_2)$ is given by
      \begin{equation}
         \cos e = \sin b_1 \sin b_2
                + \cos b_1 \cos b_2 \cos (l_1 - l_2). \label{Eq1}
      \end{equation}

      Concerning the second subproblem, different measures may be
      applied. After testing some possibilities, we decided to quantify
      the dissimilarity of two Gaussians $E_i$ and $E_j$ with the
      parameter
      \begin{equation}
         S^\prime = \frac{\int_{-\infty}^\infty (E_i - E_j)^2 dV}
                         {\int_{-\infty}^\infty E_i^2 dV +
                          \int_{-\infty}^\infty E_j^2 dV}, \label{Eq3}
      \end{equation}
      where the Gaussians $E$ are given by
      \begin{equation}
         E(x) = T \exp {\left[ - \frac{(x - V)^2} {2 W^2} \right] }, \label{Eq2}
      \end{equation}
      $T$ is the height of the component at its central velocity $V$,
      and $W$ determines the width of the Gaussian. $S^\prime$
      characterizes the squares of the deviations of one Gaussian from
      another, normalized by the sum of the squares of the deviations of
      both components from the zero line. If two Gaussians are exactly
      the same $S^\prime = 0$, and when they become more and more
      different $S^\prime \rightarrow 1$.

      After integration we may write
      \begin{equation}
         S^\prime = 1 - 2\frac{T_i W_i T_j W_j}{T_i^2 W_i + T_j^2 W_j}
                  \left( \frac{2}{W_i^2 + W_j^2} \right)^{1/2}
           \exp \left[ - \frac{(V_i - V_j )^2}{2(W_i^2 + W_j^2)} \right]. \label{Eq4}
      \end{equation}
      We can see that computationally this formula is somewhat
      inconvenient as in the most interesting region of near zero
      values, $S^\prime$ is found as a difference of two nearly equal
      quantities, which is very sensitive to roundoff errors. At the
      same time, the exact meaning of the distance parameter, used in
      computations, is actually unimportant as long as it is any
      monotonic function of a parameter, which has a meaningful
      interpretation. Therefore, for computations we may replace the
      parameter $S^\prime$ with $S = - \ln (1 - S^\prime)$, and we get
      \begin{equation}
         S = \frac{(V_i - V_j )^2}{2(W_i^2 + W_j^2)} -
 \ln \left[ 2\frac{T_i W_i T_j W_j}{T_i^2 W_i + T_j^2 W_j}
      \left( \frac{2}{W_i^2 + W_j^2} \right)^{1/2} \right]. \label{Eq5}
      \end{equation}

\begin{figure}
         \resizebox{\hsize}{!}{\includegraphics{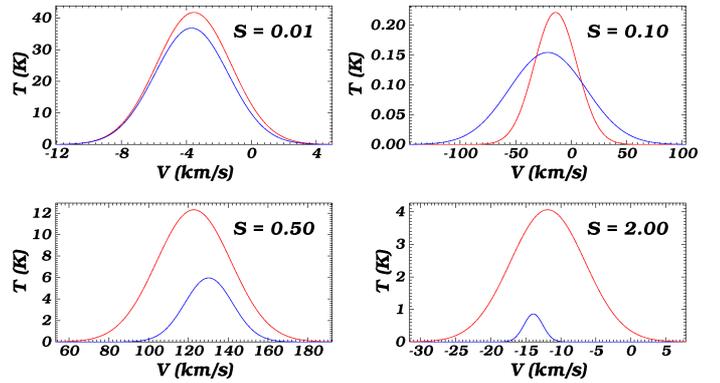}}
         \caption{Examples of pairs of Gaussians with different
            dissimilarity values $S$.}
         \label{Fig01}
      \end{figure}

      The parameter $S$, as defined by Eq.~\ref{Eq5}, compares the
      values of two Gaussian functions at all possible velocities and
      corresponds well to the natural human understanding of the
      similarity of two curves: the curves are similar when they are
      everywhere close to each other. $S = 0$, when two Gaussians have
      exactly the same values of their parameters and for increasingly
      different components $S \rightarrow \infty$. In Fig.~\ref{Fig01}
      this is illustrated by some pairs of Gaussians with different
      values of $S$.

      It is easy to see that this definition also well compensates for
      uncertainties in the determination of the Gaussian parameters, as
      discussed in Sec. 4.2. of the Paper I. For example, in the
      presence of noise our decomposition program gives the most
      unreliable values for the central velocities and widths for the
      widest Gaussians (component D in Fig. 10 of Paper I). However,
      when the line widths become larger, the differences in central
      velocities and also in widths become less important in the first
      addend of Eq.~\ref{Eq5}. Therefore, we may conclude that our
      dissimilarity measure treats the observed lines of different width
      with more or less the same precision. This is good for comparison
      of two independent Gaussians, but may pose problems for the
      clustering, as all natural gradients in parameter values become
      increasingly important for narrower components. It may turn the
      detection of small, bright, but cold clouds of \ion{H}{i} rather
      problematic. We will return to this in the next section of the
      paper, where we present the search results of the coldest clouds
      in the Galaxy.

      \subsection{Decreasing the storage needs}

      In the previous subsection, we divided the calculation of the
      ``distance'' between Gaussians into two different estimates
      (Eqs.~\ref{Eq1} and \ref{Eq5}). For general hierarchical
      clustering we must join the obtained estimates into one, but in
      our case this is not so important. The original LDS2 and IARS data
      are given in a regular grid and we are looking for continuous
      clouds. The cloud can be continuous only if it is observed at
      least in one neighboring profile of any given observation
      belonging to this cloud. However, in the regular grid we know in
      advance, which observations are the neighbors of the initial one.
      This knowledge considerably reduces the amount of required
      computations and the storage needs: there is no necessity to
      compute the distance of a given Gaussian from all others, but only
      from the components of the neighboring profiles and as the
      neighbors are on the sky at more or less equal distances, we may,
      at least in some approximation, ignore in the computation the
      spatial part of the Gaussian distances. With this simplification,
      the task is reduced to a manageable size, and most of the
      following decisions in the fine-tuning of the clustering algorithm
      were based on the comparison of the results of trial computations
      with existing cloud catalogs (mostly HVCs).

      The next step in further reduction of the storage needs was
      connected with the third problem, described at the beginning of
      this section: the avoidance of merging most Gaussians into one
      giant cloud around the galactic plane. We decided that every
      Gaussian of each profile represents a different feature in the
      real gas cloud. For example, in HVCs we often get two Gaussians in
      the same profile at nearly the same velocity, and most likely they
      both describe the same physical cloud, but the narrow Gaussian
      represents the properties of the gas in compact cold cores and the
      wider component describes the gas in the more extended warmer
      envelope of the cloud (Kalberla \& Haud \cite{Kal06}; Paper IV).
      We decided to consider such features as different entities of ISM
      and therefore to apply a restriction that every cloud of Gaussians
      may contain only one component from each profile. In this way, if
      in some problems we need to consider ``cores'' and ``envelopes''
      together, we may join corresponding clouds for this particular
      task, but if we allow them to merge from the beginning, it would
      be harder to separate different subclouds for some other studies.

\begin{figure}
         \resizebox{\hsize}{!}{\includegraphics{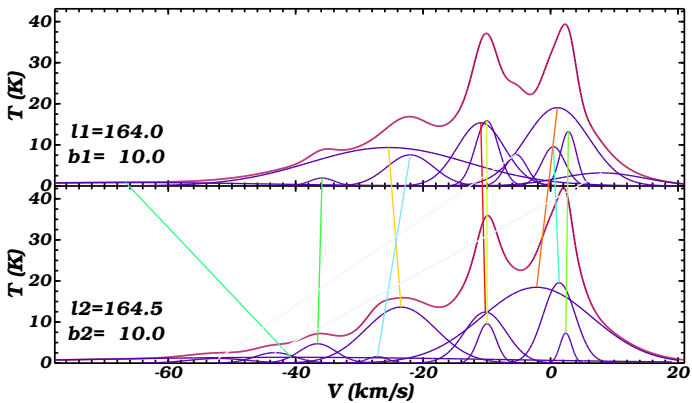}}
         \caption{Example of the pairing of Gaussians of two neighboring
            profiles. The Gaussian representations of the profiles are
            given with the raspberry lines and the individual Gaussian
            components with the violet ones. The colored straight lines
            connect the tips of the paired Gaussians. The hues of these
            lines (from red to blue) are given in the order of the
            goodness of the pair (from the best to the worst) and the
            saturations (HSV) are calculated as $2.0/(2.0+S)$. Two
            nearly invisible alice blue and magnolia lines from the
            lower left to the upper right parts of the figure connect
            the clearly unrelated Gaussians. The connections correspond
            to $S=24.0$ and $S=44.8$, and these pairs are not used in
            the clustering process. The centers of very wide components
            are sometimes located at rather different velocities, but
            they still may participate in the clustering process (spring
            green line from the lower center to the upper left,
            corresponding to $S=0.541$). A similar value of $S$ for much
            narrower Gaussians is represented by a nearly vertical lime
            green line at about $V=-36~\mathrm{km\,s}^{-1}$.}
         \label{Fig02}
      \end{figure}

      This decision has a useful side-effect, that every Gaussian in one
      profile may have only one partner in each neighboring profile, and
      we need not store the distances between all possible combinations
      of the components of two neighboring profiles, but just one
      distance per every Gaussian in the profile with the smaller number
      of components. In other words, part of the global clustering
      process may be carried out locally between each pair of profiles
      by choosing in theses profiles the best (most similar) pairs of
      Gaussians, and only the distances in these best pairs must be
      forwarded to the global merging phase of the algorithm. Such
      pairing of Gaussians of two neighboring profiles is illustrated in
      Fig.~\ref{Fig02}).

      However, even forwarding of the distances of the best pairs may be
      restricted somewhat more. Not every Gaussian in one profile can be
      successfully paired with some other in a neighboring profile.
      Some features may be present only in a single profile, or do not
      propagate into other profiles in some particular direction. Of
      course, mathematically it is possible to find a partner for every
      Gaussian of the profile, which has less components in a pair of
      profiles, but for some such pairs of Gaussians the value of $S$
      becomes very large (in Fig.~\ref{Fig02} the lines with the least
      saturated colors) and these large distances are clearly useless in
      the final clustering process. As a result, we stored the info only
      for those pairs, which had $S < 2$. Altogether we got in this way
      4\,946\,775 distances or links between Gaussians.

      \subsection{The clustering algorithm}

      The final procedure for defining clouds was as follows:
      \begin{enumerate}
         \item find for profile at any $(l, b)$ the $(l \pm 0 \fdg 5/
            \cos b, b)$, $(l, b \pm 0 \fdg 5)$ and $(l \pm 0 \fdg 5/
            \cos b, b \pm 0 \fdg 5)$ partners;
         \item for each pair of profiles calculate the dissimilarities
            (Eq.~\ref{Eq5}) of all possible pairs of Gaussians;
         \item find the most similar pairs of Gaussians for each pair of
            profiles and store the links with $S < 2$;
         \item build the index of links in the ascending order of $S$;
         \item merge the pairs of Gaussians into clouds in the order of
            the found index.
      \end{enumerate}

      At every merging step the following procedures were passed:
      \begin{enumerate}
         \item if two merging Gaussians already belonged to the same
            cloud, the corresponding link was rejected;
         \item if two merging clouds contained different Gaussians
            from the same profile, the clouds were not merged and the
            corresponding link was rejected;
         \item if the global properties of two merging clouds were too
            different, their merging was postponed;
         \item all remaining links were stored into the list of active
            links (links, which have passed all the tests and therefore
            actually merge something).
      \end{enumerate}

      The last but one procedure certainly needs some clarification. By
      using a pure single link clustering algorithm, we sometimes found
      the cases where two clouds with rather different average
      properties merged, as they touched each other at some point on
      their outer perimeter. As this was undesirable, we added a
      corresponding test and modified the algorithm somewhat in the
      direction of the global link methods. For this, when the merging
      of clouds, containing more than one Gaussian, was initiated by
      Gaussians with dissimilarity $S$, we:
      \begin{enumerate}
         \item calculated for both merging clouds their total emission
            profiles as a sum of all Gaussians in a particular cloud;
         \item calculated the dissimilarity $S_\mathrm{Cl}$ for these
            profiles. Of course, now we had to do the integration in
            Eq.~\ref{Eq2} numerically, and then apply the transform to
            get the result similar to $S$ from Eq.~\ref{Eq5} for single
            Gaussians;
         \item made the decision about merging the clouds:
            \begin{itemize}
               \item if $S_\mathrm{Cl} \le S$, corresponding clouds were
                  merged and link $S$ was written to the list of active
                  links;
               \item if $S < S_\mathrm{Cl} < 2$,
                  \begin{enumerate}
                     \item the clouds were not merged,
                     \item the value of $S$ was replaced with
                        $S_\mathrm{Cl}$,
                     \item the original link was deleted from the
                        ordered list of links,
                     \item a new link was generated in the list of links
                        in the place, corresponding to the value of
                        $S_\mathrm{Cl}$,
                     \item when some time later the clustering process
                        reached that new link, it was treated as all
                        other links between single Gaussians;
                  \end{enumerate}
               \item if $S_\mathrm{Cl} \ge 2$, the clouds were not merged
                  and the corresponding link was rejected.
            \end{itemize}
      \end{enumerate}

      It may seem that the added procedure is rather limiting and
      determines the whole clustering process, but the actual tests did
      not confirm this. With merging the Gaussians, the widths of the
      resulting cloud profiles initially grow rapidly and the comparison
      of such wide profiles gives rather small values of
      $S_\mathrm{Cl}$. As a result, at the beginning of the merging
      process the comparison of global properties of the merging clouds
      rarely changes the run of clustering. It becomes more important at
      later stages when the Gaussians with larger mutual differences are
      initiating the merging. Therefore, the outcome of the clustering
      process has not changed dramatically, but nevertheless the results
      are brought into better accordance with those, obtained from human
      inspection of the data.

      As a result of the described algorithm, we obtained a list of
      1\,350\,655 active links between Gaussians. It represents the
      clustering dendrogram and for getting a list of clouds or
      clusters, we must cut this dendrogram at an appropriate level.

   \section{Very cold clouds}

      \subsection{Finding the clouds}

      In the previous section, we discussed that the proposed cloud
      compilation algorithm is expected to work better with relatively
      wide Gaussians, but there may be certain problems with finding the
      coldest clouds of very narrow Gaussians. First tests in the width
      range, corresponding to HVCs, have shown that the results are
      acceptable, but we will return to this in our next papers. Here we
      would like to present the results for the worst case of very
      cold clouds.

      \begin{figure}
         \resizebox{\hsize}{!}{\includegraphics{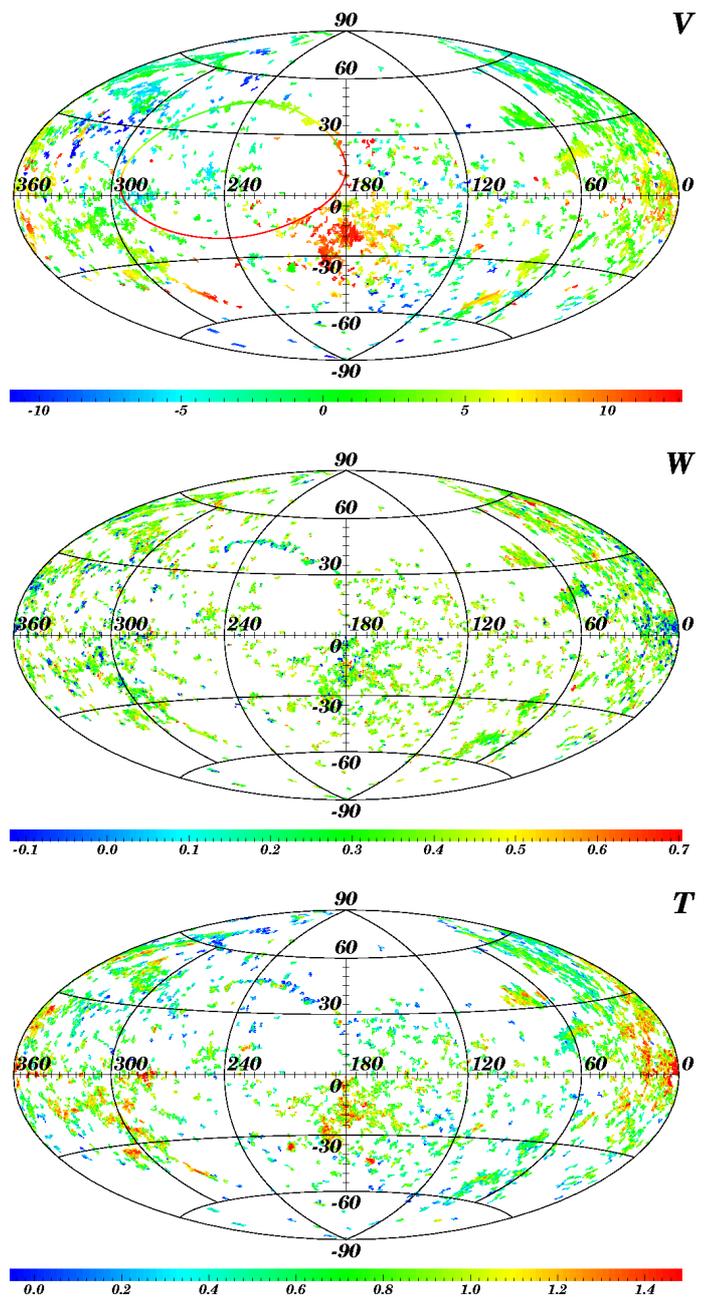}}
         \caption{The velocities, line-widths and brightness
            temperatures in clouds of at least 7 Gaussians, compiled by
            our clustering algorithm. Shown are the objects for which
            the mean Gaussian height is $\ge 1.0~\mathrm{K}$, $FWHM \le
            3.0~\mathrm{km\,s}^{-1}$, $|V| \le 15~\mathrm{km\,s}^{-1}$
            and the parameters of at least half of the Gaussians satisfy
            Eq. 4 of Paper II (components are likely not radio
            interferences). The color scales are for $V$, $\lg W$ and
            $\lg T$, respectively. The color line in the upper panel
            represents the sky positions and the velocities of our model
            ring (see Sec. 3.2).}
         \label{Fig03}
      \end{figure}

      To our knowledge, the coldest clouds in the Galaxy, found so
      far primarily in \ion{H}{i} emission, are the two gas
      concentrations at near-zero velocity around $(l = 225\degr, b =
      +44\degr)$ and $(l = 236\degr, b = +45\degr)$, discovered by
      Verschuur (\cite{Ver69}), and afterwards studied in more details
      by Verschuur \& Knapp (\cite{Ver71}) and Knapp \& Verschuur
      (\cite{Kna72}). Later the third cloud around $(l = 213\degr, b =
      +41\degr)$ was added to the first two by Heils \& Troland
      (\cite{Hei03}), and one of the recent detailed studies of these
      clouds is that by Meyer et al. (\cite{Mey06}). They observed the
      interstellar \ion{Na}{i} D1 and D2 absorption toward 33 stars,
      derived a cloud temperature of $20_{-8}^{+6}~\mathrm{K}$ and
      placed a firm upper limit of $45~\mathrm{pc}$ on the distance of
      the clouds. This distance corresponds to the upper limit of
      the linear size of the clouds of about $5~\mathrm{pc}$. Redfield
      \& Linsky (\cite{Red08}) have interpreted these clouds as a result
      of the collision of warm high-velocity Gem Cloud with the slower
      moving Leo, Aur, and LIC Clouds in the Local Interstellar Medium
      (LISM). At the same time, the properties of the clouds seem
      to be also rather similar to the temperatures (mostly $10 < T_S <
      40~\mathrm{K}$) and dimensions (in parsec scale) of numerous
      \ion{H}{i} self-absorption features, found near the galactic plane
      (e.g. Gibson et al. \cite{Gib00}; Dickey et al. \cite{Dic03};
      Kavars \cite{Kav05}; Hosokawa \& Inutsuka \cite{Hos07}). Our
      interest in the subject is to test if the clustering algorithm
      finds these clouds and in the case of a positive answer, to look
      for similar features all over the sky.

      For this test we first constructed the dendrogram for all
      Gaussians in our decomposition and inspected the resulting clouds
      for different values of the cutting level of the dendrogram. From
      this inspection we chose for the final cutting level the value
      $S_\mathrm{Cu} = 0.44$. In this way we obtained 94\,874 clusters
      of Gaussians and 236\,306 components remained detached from the
      others. The largest cloud (12\,585 Gaussians) in the obtained list
      corresponds to relatively smooth warm neutral medium at high
      galactic latitudes, but the list contains also many very small
      clouds of 2-3 Gaussians in each (the cluster size distribution
      follows the power law with the slop of about 1.9). As Verschuur \&
      Knapp (\cite{Ver71}) have estimated that their cool clouds have
      diameters of at least $1\fdg5$, we are not interested in the
      smallest clouds in our list, and in the following, we will
      consider only the clouds, containing at least 7 Gaussians (in LAB
      one profile represents an area of 0.25 square degrees and 7
      profiles cover the area, corresponding to the cloud with the
      diameter of $1\fdg5$). In our list, there are 21\,224 clouds of
      such size.

      To search for the coldest clouds in the list, we must apply some
      additional selection criteria. First of all, we are looking for
      clouds, consisting of relatively narrow Gaussians. In Paper III,
      we demonstrated that the mean line-width of the \ion{H}{i} 21-cm
      radio lines of the cold neutral medium of our Galaxy is $FWHM =
      3.9 \pm 0.6~\mathrm{km\,s}^{-1}$. Therefore, the gas, having $FWHM
      \le 3.0~\mathrm{km\,s}^{-1}$, may be considered already as a very
      cold gas and we will look for the clouds for which the mean width
      of the Gaussians is below this limit. In Paper II, we also
      demonstrated that many weak and/or very narrow Gaussians do
      not represent the actual \ion{H}{i} emission of the Galaxy, but
      are more likely due to observational noise or radio interferences.
      Here we are not interested in these Gaussians, and therefore we
      apply the selection criteria, given by Eqs. 4 and 5 of Paper II.
      However, now we do not apply these criteria to single Gaussians,
      but to the clouds obtained from our clustering process.

      From Eqs. 4 and 5 of Paper II, it follows that the narrowest
      Gaussians, which most likely represent the galactic \ion{H}{i},
      have the heights $\ga 0.95~\mathrm{K}$. Therefore, we consider
      only these clouds, for which the mean height of their Gaussians is
      $\ge 1.0~\mathrm{K}$. At first sight, a similar selection ($FWHM
      \ge 1.25~\mathrm{km\,s}^{-1}$, corresponding to Eq. 4 of Paper II)
      may also be applied to the width of the Gaussians. However, we are
      looking for clouds with the narrowest Gaussians, and some of the
      real lines may be even narrower than interferences with $T \ge
      1.0~\mathrm{K}$. Therefore, as such selection may reject not only
      the interferences, but also a considerable amount of Gaussians of
      main interest in our study, this selection cannot be applied
      directly. At the same time, the selection rule given by Eq. 4 of
      Paper II, applies only statistically and it turned out that better
      results can be obtained by rejecting the clouds, for which more
      than half of their Gaussians do not satisfy Eq. 4 of Paper II.
      Nevertheless, some confusion with the interferences still remains.

      \begin{figure*}
         \resizebox{\hsize}{!}{\includegraphics[width=17cm]{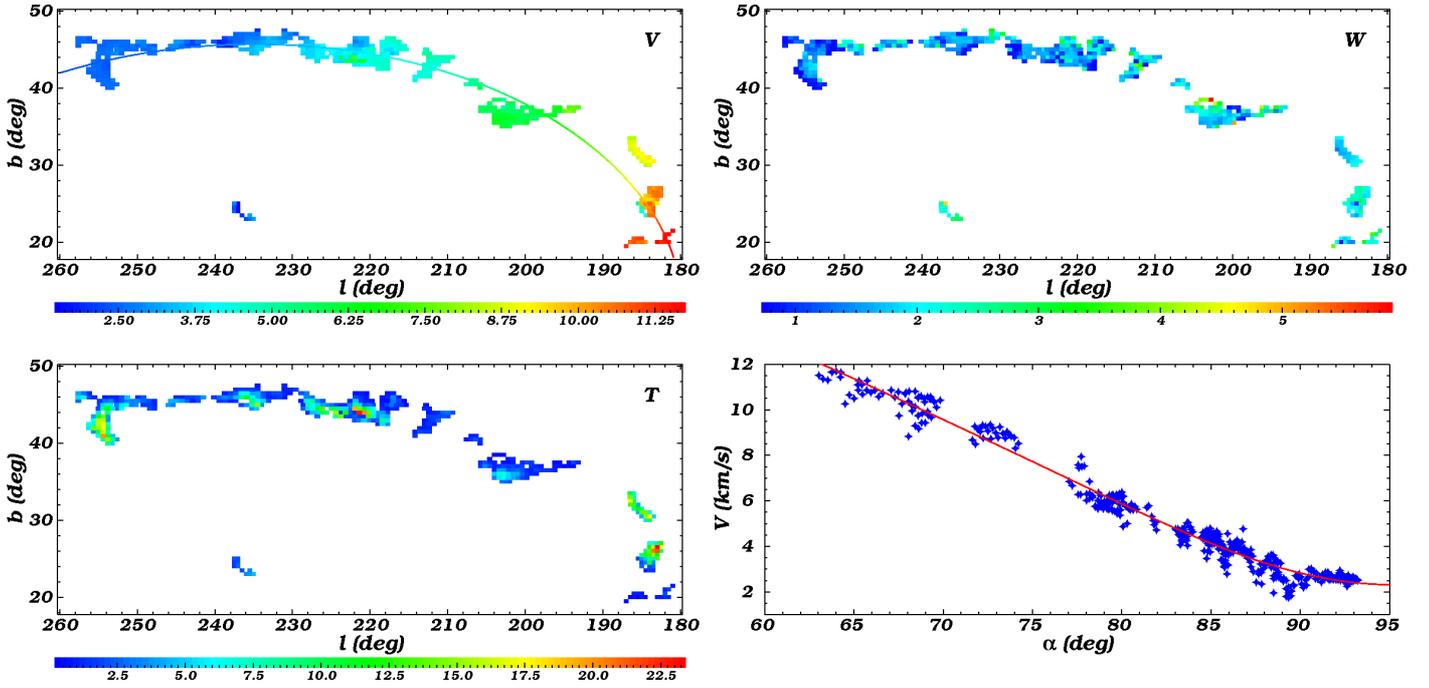}}
         \caption{The velocities, line-widths and brightness
            temperatures of the clouds in the region of the observable
            part of the ring. Shown are the objects, represented by at
            least 7 Gaussians, for which the mean Gaussian width $FWHM
            \le 2.7~\mathrm{km\,s}^{-1}$, $2.25 \le V \le
            11.55~\mathrm{km\,s}^{-1}$ and the parameters of at least
            half of the Gaussians satisfy Eq. 4 of Paper II (components
            are likely not radio interferences). The color scales are
            for linear values of $V$, $W$ and $T$. The color line in the
            upper left panel represents the sky positions and the
            velocities of our model ring. A more detailed comparison of
            the observed LSR velocities of the ring clouds (blue
            diamonds) with the model velocities (red line) is given in
            the lower right panel of the figure. Here the abscissa is a
            polar angle of the ring point in the ring plane.}
         \label{Fig04}
      \end{figure*}

      After applying all the described selection criteria we received a
      list of 1\,380 cold clouds. However, when looking at these clouds
      we saw that the clouds with the highest velocities (concentrated
      around $+50$ and $+100~\mathrm{km\,s}^{-1}$) were located only in
      a very narrow band around the galactic plane (all at $|b| <
      22\degr$, most at $|b| < 5\degr$). We have stressed several times
      that in these regions the Gaussian decomposition gives relatively
      unreliable results, and the corresponding Gaussians are with high
      probability not directly related to the physical properties of the
      ISM. Therefore, we decided to reject also these clouds by applying
      the requirement $|V| \le 15~\mathrm{km\,s}^{-1}$ on the mean
      velocities of the clouds. In this way, we rejected 44 more small
      clouds. All remaining clouds are presented in Fig.~\ref{Fig03}.

      When applying the described selection criteria on clusters,
      obtained with different values of the cutting level,
      $S_\mathrm{Cu}$, of the dendrogram, we found that for $0 <
      S_\mathrm{Cu} < 0.27$ the number of Gaussians in the selected
      clouds increases rapidly. For $0.27 \ge S_\mathrm{Cu} < 0.75$, the
      pictures similar to Fig.~\ref{Fig03} remain nearly unchanged with
      only a slight maximum in the number of Gaussians for
      $S_\mathrm{Cu} = 0.44$. After $S_\mathrm{Cu} = 0.75$ the number of
      Gaussians starts to decrease as gradually wider and wider
      Gaussians are linked to the existing clouds and the average
      line-widths of clouds grow above our selection limit. For
      Fig.~\ref{Fig03}, we chose the value of $S_\mathrm{Cu}$, which
      gave the highest number of Gaussians in the selected clouds.

      \subsection{Verschuur's clouds}

      From Fig.~\ref{Fig03}, we can see that the cold clouds around $(l
      = 225\degr, b = +44\degr)$, $(l = 236\degr, b = +45\degr)$ and $(l
      = 213\degr, b = +41\degr)$, mentioned at the beginning of the
      previous subsection, are clearly visible. Moreover, in this figure
      these clouds seem to be a part of a more extended narrow string of
      clouds, covering on the sky about $80\degr$ from $(l \approx
      181\degr, b \approx +20\degr)$ to $(l \approx 258\degr, b \approx
      +47\degr)$. This is in good agreement with the remark by Heils \&
      Troland (\cite{Hei03}), who mentioned that other narrow \ion{H}{i}
      21-cm emission lines could be found in the extended region around
      the clouds, studied in their paper. Nevertheless, they limited
      their interest only to the longitude interval of $200\degr \le l
      \le 240\degr$ and reported the broken ribbon of cold \ion{H}{i}
      gas stretching over $20\degr$ across the constellation Leo.

      In more details, these clouds are plotted in Fig.~\ref{Fig04}. To
      better separate them from other features in the same sky region,
      we have used here even more severe selection criteria ($FWHM \le
      2.7~\mathrm{km\,s}^{-1}$, $2.25 \le V \le
      11.55~\mathrm{km\,s}^{-1}$), compared to those for
      Fig.~\ref{Fig03}. However, to increase the sensitivity and as the
      noise Gaussians are rather effectively removed from the data also
      by using only clouds of 7 or more Gaussians, we have dropped here
      the requirement $T \ge 1.0~\mathrm{K}$. Due to the changes in
      selection criteria, the considerations, similar to those used for
      selecting the dendrogram cutting level $S_\mathrm{Cu} = 0.44$ for
      Fig.~\ref{Fig03}, have lead in this case also to somewhat higher
      $S_\mathrm{Cu} = 0.53$. As a result, we have obtained a chain of
      clouds, which seems to follow some arc, well populated in its
      higher galactic longitude half and more opened at lower
      longitudes.

      We can see a clear velocity gradient along this ribbon of clouds
      with the average velocities of the clouds increasing by about
      $9~\mathrm{km\,s}^{-1}$ per length of the arc. A similar, but a
      much weaker gradient holds also for the line-widths: the average
      $FWHM$ of the clouds increases by $1~\mathrm{km\,s}^{-1}$ from the
      higher galactic longitudes towards the lower longitudes. It also
      appears that the clouds tend to be brighter near their centers
      than in outer regions, which is an expected behavior for real gas
      clouds. In this way, while also the lower longitude part of the
      string of clouds is interrupted in some places by voids, the
      coherence of its characteristics strongly indicates that it is
      really the same physical feature. It remains questionable, whether
      the clouds at $(l = 237\degr, b = 24\degr, V =
      2.3~\mathrm{km\,s}^{-1})$ and $(l = 184\degr, b = 25\degr, V =
      4.5~\mathrm{km\,s}^{-1})$ also belong to the same structure as
      they deviate from the others considerably on the sky or in
      velocity. Therefore, we have not used them in the following
      discussion.

      The smoothness of the observed structure was rather tempting for
      its modeling. As arced shapes often hint at circular structures,
      seen under some angle to their plane, we decided to model this
      string of clouds and their velocities as a partial gas ring
      somewhere in space, which may move relative the Local Standard of
      Rest (LSR) as a whole and also rotate around its center and expand
      away from this center. We found that such model very well
      describes both, the apparent location of the clouds on the sky and
      their observed velocities. According to the obtained model, the
      center of the ring is located in the direction $(l = 236\fdg2 \pm
      0\fdg9, b = -13\fdg2 \pm 0\fdg3)$, its apparent major axis is
      inclined by $4\fdg7 \pm 0\fdg9$ to the galactic plane and the
      angle between the ring plane and the line of sight to its center
      is $14\fdg5 \pm 0\fdg3$. The radius of the ring is seen under the
      angle of $41\fdg8 \pm 0\fdg3$ along the apparent major axis of the
      observed structure. The ring as a whole moves with the velocity of
      $21.0 \pm 0.5~\mathrm{km\,s}^{-1}$ in the direction $(l = 193\fdg3
      \pm 1\fdg2, b = 2\fdg2 \pm 0\fdg7)$, rotates clockwise with the
      velocity of $10.6 \pm 0.4~\mathrm{km\,s}^{-1}$ and its expansion
      speed is $26.2 \pm 0.7~\mathrm{km\,s}^{-1}$. As errors of these
      parameters are given the 99.73\% confidence limits obtained from
      the bootstrapping.

      The projection of the model ring onto the sky is shown with a line
      in the first ($V$) panels of Figs.~\ref{Fig03} and \ref{Fig04}.
      The color of the line corresponds to the line of sight velocity of
      each ring point. The fit of the model to the observed gas
      velocities is shown in the lower right part of Fig.~\ref{Fig04}.
      From Fig.~\ref{Fig03}, we can see that actually the same structure
      seems to continue even beyond the lower latitude border of
      Fig.~\ref{Fig04}, and it can be followed down to about $(l =
      225\degr, b = -19\degr)$. However, this continuation of the ring
      is rather sparsely populated with relatively small clouds and is
      located near the galactic plane, where the Gaussian decomposition
      cannot be considered to be reliable. Therefore, we will not
      discuss this continuation in greater detail than just mentioning
      that when the parameters of the ring were estimated only from a
      $30\degr$ segment of the whole ring (as seen from the ring center
      and indicated in the lower right panel of Fig.~\ref{Fig04}), in
      total the visible part of the ring may extend to nearly half
      ($162\degr$) of the full circle. For the other half there seems to
      be no good candidates for the same structure. But, of course, also
      the location of the model ring is rather uncertain in these
      regions.

      \subsection{The ring in space}

      In most \ion{H}{i} profiles the emission, corresponding to the
      clouds under discussion, appears as a very narrow and relatively
      strong line, not seriously blended by a broader-velocity,
      lower-intensity emission component. Therefore, it may seem to be
      easy to derive from the parameters of our Gaussians some estimates
      for the physical conditions inside these clouds. Unfortunately, as
      briefly mentioned in the Introduction, this is not true. Already
      Verschuur \& Knapp (\cite{Ver71}) demonstrated that the shapes of
      these narrow emission lines are actually not Gaussians, but they
      are considerably influenced by saturation. They derived the spin
      temperature by assuming the optical depth to be a Gaussian
      function of the frequency and fitting the observational data to
      the equation of transfer. We cannot use even this path, as the
      velocity resolution of the LAB data is more than 10 times lower
      than that of the data used by Verschuur \& Knapp (\cite{Ver71})
      and therefore the actual line-shapes are mostly unresolved.

      Nevertheless, we decided to take a further step and to obtain at
      least some preliminary estimate for the distance of the ring. In
      doing so, we followed the procedure described by Haud
      (\cite{Hau90}). These estimates are based on the assumption that a
      correlation exists in cold \ion{H}{i} clouds between the clouds
      internal velocity dispersion and its linear dimensions, similar to
      the one observed for molecular clouds (Larson \cite{Lar81} and
      many others since then). We will not discuss here all the
      questions, related to the existence or meaning of such
      correlation, but use it just as a possible tool, which may or may
      not give some acceptable results. We followed exactly the same
      procedure as described in Haud (\cite{Hau90}) with the only
      exception that we did not correct the LSR velocities of the clouds
      to the Galactic Standard of Rest (GSR), as in this case it is most
      likely unjustified. Instead, we removed the large scale velocity
      gradients and projection effects in the ribbon clouds using our
      ring model. In this way, we obtained the distance estimates for
      all ring clouds and using our model of the ring, converted them to
      estimates of the distance of the ring center.

      As expected, we got rather scattered results, but in general, the
      distance estimates of individual clouds agreed with our ring
      model, which indicates that the lower longitude tip of the band of
      clouds is located from us about twice as far as the higher
      longitude tip. As the scatter of the obtained estimates was
      considerably higher for estimates, based on smaller (covering
      fewer gridpoints of LDS observations) clouds than for those, based
      on larger clouds, we decided to accept for the distance of the
      ring center the weighted average of all determinations, and to use
      as weights the number of Gaussians in each cloud. In this way, we
      obtained a distance estimate of $126 \pm 82~\mathrm{pc}$. The
      error estimate corresponds only to the scatter of individual
      distance estimates and does not account for uncertainties in the
      ring model or in the method, used for obtaining these distances.
      To this distance corresponds the linear radius of the model ring
      of $113~\mathrm{pc}$ and the distance of the Verschuur's cloud A
      of $34~\mathrm{pc}$, which is in good agreement with the upper
      distance limit ($45~\mathrm{pc}$) for this cloud, established by
      Meyer et al. (\cite{Mey06}).

      Of course, with such a model a number of questions remain. First
      of all, why to model this structure as a planar ring of gas
      clouds, when most processes, which may give the expansion
      velocities, obtained for this ring, have more likely spherical
      symmetry? Moreover, when the ribbon of gas clouds covers nearly
      $80\degr$ in the galactic longitude, in our model this corresponds
      only to $30\degr$ along model ring itself. This means that
      actually we do not know anything about most of the ring, and
      therefore its parameters may contain large systematic errors.
      Also, the distance estimates are based on rather arbitrary
      assumptions, they are quite uncertain, and as the Gaussians most
      likely overestimate the widths of the actual underlying \ion{H}{i}
      lines, they must be considered as upper limits for corresponding
      actual distances.

      Nevertheless, even such model well demonstrates the coherence of
      the observed clouds, it seems to give some indication of possible
      continuation of the structure even beyond the region studied here,
      and it is interesting to see that the distance estimates of the
      individual clouds and the ring model are in general agreement with
      respect to the orientation of the gas band in 3-dimensional space.
      Moreover, we have seen that the observed behavior of the gas
      stream at lower longitudes may be understood in the framework of
      this model -- in these regions the distance of the clouds from the
      Sun increases and therefore they become apparently smaller. As we
      have selected from our clustering results only relatively large
      clouds (at least 7 Gaussians in each cloud), we may lose most of
      these apparently smaller ones from our view. Therefore, beyond
      about $(l = 182\degr, b = 20\degr)$ and the distance
      $64~\mathrm{pc}$ the stream becomes rather fragmentary and we can
      observe only some seemingly small clouds, which actually may have
      relatively large linear dimensions and line-widths. As here we
      cannot see any more the really smallest and coldest clouds, this
      may explain the increase of the average observable line-widths in
      this region. But why then the stream so abruptly terminates at its
      other end? This happens practically at the nearest point of the
      ring to the Sun.

      Wolleben (\cite{Wol07}) has proposed a model for the North Polar
      Spur (NPS) region. This model explains the results of the Dominion
      Radio Astrophysical Observatory Low-Resolution Polarization Survey
      (Wolleben et al. \cite{Wol06}), and the model consists of two
      synchrotron-emitting shells, S1 and S2 (Fig.~\ref{Fig05}). The
      same model shells were used by Frisch (\cite{Fri08}) to explain
      the higher column densities of \ion{Ca}{ii} in the galactic
      quadrants $l > 180\degr$ versus $l < 180\degr$. We studied the
      mutual placement of these shells and our ring, and found that the
      ring intersects with the shell S2 in the direction $(l = 256\degr,
      b = 43\degr)$ at the distance of $33~\mathrm{pc}$ from the Sun.
      This position exactly matches the beginning of our band of clouds,
      and the result is nearly independent of the rather indefinite
      determination of the linear size of the ring. Further to the
      higher longitudes the ring continues inside the S2 and probably
      the ring clouds are destroyed by the shell. The ring leaves the
      shell at $(l = 295\degr, b = 4\degr)$ at the distance of
      $85~\mathrm{pc}$ from the Sun. As with the lower longitude end of
      the gas stream, we may expect that at these distances the ring
      clouds, even if they exist there, are mostly unobservable.

      A problem with this explanation of the observability of the ring
      clouds is the fact that the ring intersects the S1 shell at $(l =
      235\degr$, $b = 46\degr$, $d = 33~\mathrm{pc})$, but is still well
      observable to both sides of this point. Maybe only a slight
      disturbance of the velocities of the ring clouds can be seen in
      this region. However, the explanation of the different behavior of
      the ring clouds at the intersections with two different shells may
      lie in the different ages of these shells. According to Wolleben
      (\cite{Wol07}), the S1 shell is about 6 million years old and
      observable only as a small part of ``New-Loop'' in the Southern
      galactic hemisphere. The S2 shell is 1-2 million years old and
      seems to be much more active as it is responsible for the well
      known NPS. Therefore, we may expect that the S1 shell is not any
      more energetic enough to destroy the ring clouds, as they are
      destroyed by S2. By arbitrarily using the standard model for the
      kinematic age of stellar wind bubbles (Weaver et al.
      \cite{Wea77}), we may estimate that the age of the ring itself is
      about 2.5 million years, quite comparable with the age of the S2
      shell. However, the physical mechanisms responsible for the
      production of such cold clouds in the environment of the hot Local
      Bubble (LB) are still poorly understood (Stanimirovi\'{c}
      \cite{Sta09}).

      \subsection{Other cold clouds}

      In Fig.~\ref{Fig03}, we can also see some other concentrations of
      the gas besides the one described in previous subsections. Many of
      them are even more prominent than the narrow band of clouds, we
      have just discussed. Of those we can mention, for example, four
      clouds at approximately $l = 70\degr$ and $b = -50\degr$,
      $-30\degr$, $15\degr$, $35\degr$, but also the wide bands of
      clouds from $(l = 10\degr, b = 15\degr)$ to $(l = 290\degr, b =
      70\degr)$, from $(l = 30\degr, b = 15\degr)$ to $(l = 0\degr, b =
      70\degr)$ and from $(l = 320\degr, b = -10\degr)$ to $(l =
      260\degr, b = -55\degr)$. Some very narrow Gaussians, seen in the
      IARS part of the LAB still represent radio interferences, and we
      will not discuss the concentrations near the galactic plane.

      However, most of these gas concentrations differ from ring clouds
      in many respects. First of all, the average width of corresponding
      Gaussians is somewhat (up to 1.5 times) larger than that of the
      ring clouds. When in ring clouds the narrowest Gaussians form
      small knots in the environment of somewhat wider Gaussians, in
      many places of the concentrations, mentioned in this subsection,
      the situation seems to be quite opposite -- small knots of wider
      Gaussians are surrounded by not so wide ones. The velocity
      distribution is also different. For the ring clouds, we may
      observe a very smooth velocity variation along the band of clouds
      and no velocity variations of the same magnitude inside individual
      clouds. For other gas concentrations, general velocity gradients
      are less prominent, but there are considerable velocity variations
      in the smaller regions. The only exception may be the cloud
      complex from $(l = 320\degr, b = -10\degr)$ to $(l = 260\degr, b =
      -55\degr)$, where the clouds further from the galactic plane have
      clearly higher velocities than those more close to the galactic
      plane.

      \begin{figure}
         \resizebox{\hsize}{!}{\includegraphics{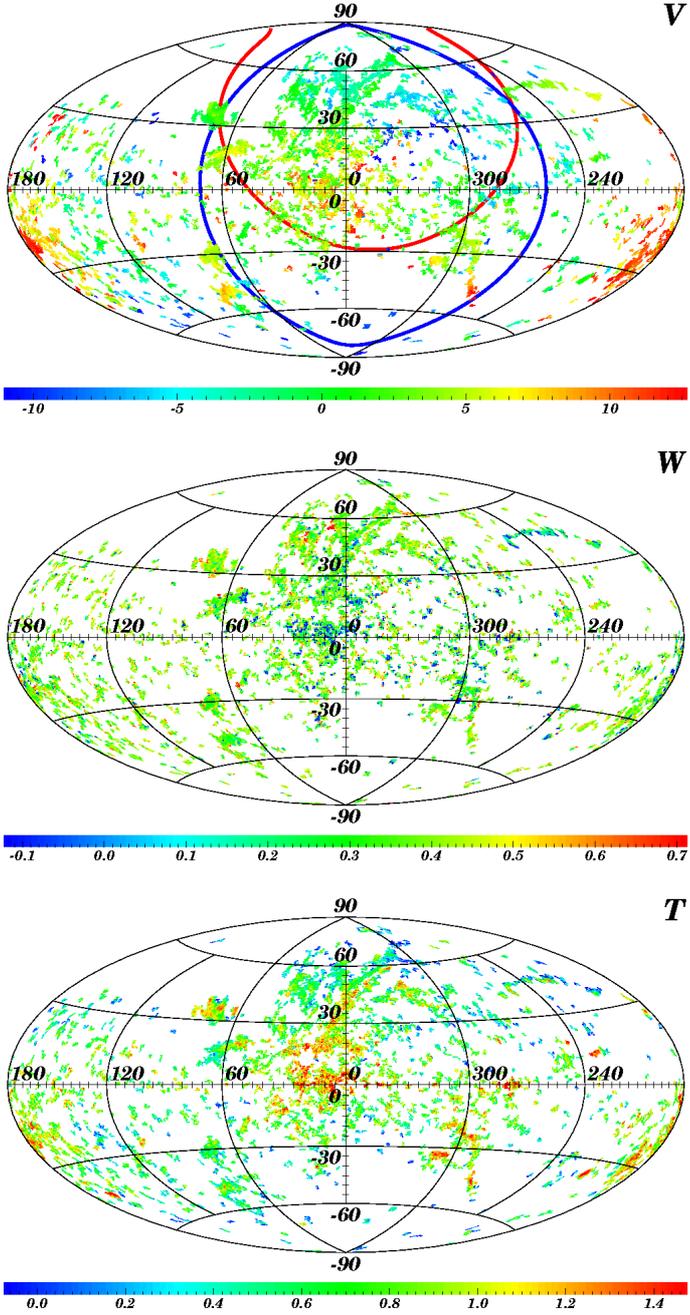}}
         \caption{The same as Fig.~\ref{Fig03}, but now centered on
            the galactic center and without the data about the ring
            model. Instead, the S1 and S2 shells are marked with thick
            background lines in the $V$panel. The red curve gives the
            visible outer boundary of the S2, but as we are located
            inside the S1, the blue curve only approximately outlines
            the region, containing the best visible part of S1.}
         \label{Fig05}
      \end{figure}

      \begin{figure}
         \resizebox{\hsize}{!}{\includegraphics{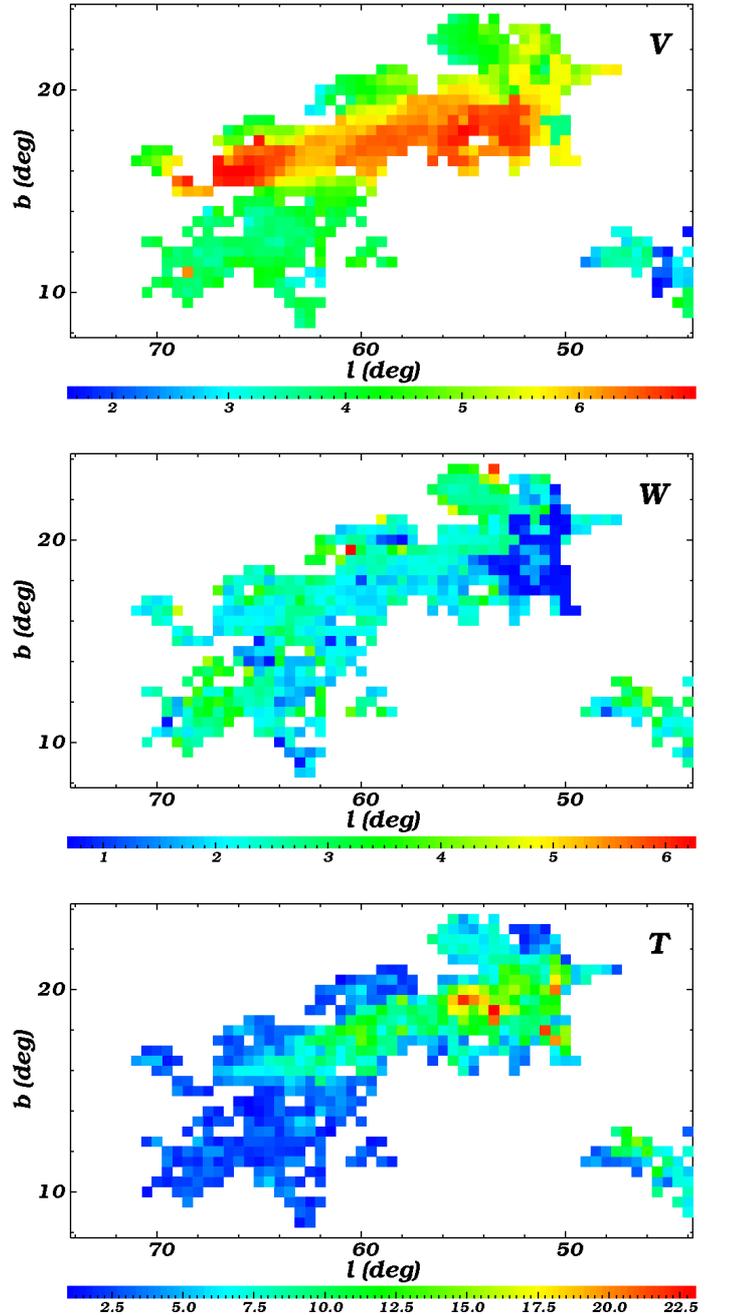}}
         \caption{The same as Fig.~\ref{Fig04}, but for clouds around $l
            = 60\degr, b = 15\degr$ and without the data about the ring
            model.}
         \label{Fig06}
      \end{figure}

      The location of the ring clouds versus others on the sky is also
      considerably different. We demonstrated that the ring extends from
      the surface of S2 shell away towards the galactic anticenter
      region. We have not modeled the spatial locations of the other
      cloud complexes, but in projection onto the sky most of them seem
      more likely to follow the shapes of the best observable parts of
      the shells S1 and S2 than to intersect the surfaces of these
      shells (Fig.~\ref{Fig05}). The clouds from $(l = 10\degr, b =
      15\degr)$ to $(l = 290\degr, b = 70\degr)$ and from $(l = 30\degr,
      b = 15\degr)$ to $(l = 0\degr, b = 70\degr)$ project onto the S2
      shell and in their general shape seem to follow the regions, where
      the line of sight is tangent to shell S2, and therefore they may
      be related to the NPS.

      The clouds at $l = 70\degr$ and $b = -50\degr$, $-30\degr$,
      $15\degr$, $35\degr$ rather exactly border the best observable
      part of the S1 and may be therefore somehow related to this shell.
      Of these four clouds the one at about $l = 60\degr, b = 15\degr$
      extends in sky projection deepest into the shell S1 and maybe even
      into S2. It is interesting that also the internal structure of
      this cloud seems to hint at the interaction with external media
      (Fig.~\ref{Fig06}). Its densest parts have the highest velocities
      and resemble the stream of gas, nearly perpendicular to the
      borders of the images of the S1 and S2 shells. This relatively
      dense and fast-moving gas is mostly surrounded by the envelope
      with smaller observed line of sight velocities and surface
      densities. In front of the head of this comet-like stream (in the
      region, closest to the S1 and S2) is located the most narrow-lined
      (the coolest) gas. Other three clouds, which seem to be located
      further away from the shells surfaces, do not have such
      distinctive cometary structures. However, we must admit that also
      Fig.~\ref{Fig06} changes, when we relax our selection criteria on
      line widths. In this case, the stream obtains an extension with
      clouds of slightly wider lines in front of the "comet's" head. At
      the same time, as these additional clouds have smaller sizes and
      higher line widths, in space they may be located somewhere behind
      the "comet", illustrated in Fig.~\ref{Fig06}.

      The gas from $(l = 320\degr, b = -10\degr)$ to $(l = 260\degr, b =
      -55\degr)$ once again seems to resemble the ring clouds. The
      stream projects onto the shell S1 and its location relative to S2
      is very similar to that of the ring clouds: the stream starts near
      the outer boundary of the shell and extends nearly perpendicularly
      away from S2. Both the ring clouds and the stream described here
      have their smallest line of sight velocities near S2 and the
      velocities increase when moving away from the shell. At the same
      time, this Southern stream has much more un-ordered appearance
      than the ring clouds in the Norther sky. Nevertheless, there may
      be even a possibility to imagine that these Southern clouds form a
      part of our ring, but in this case the estimates for the ring
      parameters must be rather inaccurate. Actually, if this is the
      case, the real shape of the structure must considerably deviate
      from the perfect ring as it seems to be impossible to get an
      acceptable fit of the location and velocities of the Northern part
      of the ring when its Southern part is forced to follow the gas at
      $(l = 320\degr, b = -10\degr)$ to $(l = 260\degr, b = -55\degr)$.

   \section{Conclusions}

      So far we have decomposed the LAB database of \ion{H}{i} 21-cm
      line profiles into the Gaussian components (Paper I) and studied
      the statistical distributions of the obtained components (Papers
      II -- IV). These distributions have revealed several interesting
      structures, but have given only the probabilities with which some
      particular Gaussian may belong to one or another structure. In
      this paper, we proposed an algorithm for grouping similar
      Gaussians. In this way, we free ourselves from the need to study
      each Gaussian separately, and we may expect that all the Gaussians
      of the ``cloud'' of similar components have the same nature. It
      may also be possible to obtain some additional physical
      information from the shapes and sizes of such clouds.

      As a test problem, we have considered the separation of clouds of
      the narrowest Gaussians as on the basis of the preliminary
      considerations just this may be the hardest problem for our
      algorithm. We have demonstrated that actually the algorithm easily
      found the coldest known \ion{H}{i} clouds discovered decades ago
      by Verschuur (\cite{Ver69}). As expected, the tests indicated
      that, depending on the cutting level of the clustering dendrogram,
      our approach may divide some larger clouds into separate, most
      coherent substructures, but hopefully mostly avoids the merging of
      unrelated features. Such behavior was intentional, as it seems
      more appropriate to study a larger number of clouds of which
      everyone represents a certain type of line features than to have a
      smaller number of clouds which may mix Gaussians of different
      nature into one.

      We also found that Verschuur's clouds form only a small part of a
      much longer ribbon of presumably very cold clouds covering on the
      sky about $80\degr$. As the gas velocities and line widths vary
      along this ribbon rather smoothly, we decided to model the whole
      structure as a part of a planar gas ring which may move in space
      as a whole and also rotate around and expand away from its center.
      Such a model very well represented the observed properties of the
      gas stream and indicated that the ring center must be located at a
      distance of $126 \pm 82~\mathrm{pc}$ from the Sun in the direction
      $(l = 236\fdg2 \pm 0\fdg9, b = -13\fdg2 \pm 0\fdg3)$. The ring
      radius is about $113~\mathrm{pc}$, its apparent major axis is
      inclined by $4\fdg7 \pm 0\fdg9$ to the galactic plane and the
      angle between the ring plane and the line of sight to its center
      is $14\fdg5 \pm 0\fdg3$. The ring as a whole moves with the
      velocity of $21.0 \pm 0.5~\mathrm{km\,s}^{-1}$ in the direction
      $(l = 193\fdg3 \pm 1\fdg2, b = 2\fdg2 \pm 0\fdg7)$, it rotates
      clockwise with the velocity of $10.6 \pm 0.4~\mathrm{km\,s}^{-1}$
      and its expansion speed is $26.2 \pm 0.7~\mathrm{km\,s}^{-1}$. In
      the framework of such model the apparent gradual weakening of the
      ring clouds at the lower longitude tip of the stream is explained
      by increasing distances between the Sun and the ring clouds in
      this region, and the abrupt end of the stream at the higher
      longitude part is caused by the intersection of the ring with the
      S2 supernova shell from the model by Wolleben (\cite{Wol07}).

      We have briefly discussed also other narrow-lined \ion{H}{i}
      clouds, found by our clustering algorithm. In many respects most
      of them are somewhat different from the ring clouds (lines are
      slightly wider, velocities less coherent over the structures
      etc.). We have not attempted to model these features, but have
      noted that their locations on the sky may hint at their relation
      to supernova activities in the solar neighborhood. Anyway, as the
      line-widths of these clouds are also small, they must be
      relatively cold clouds and therefore not very large spatially. As
      these clouds of presumably small linear dimensions cover rather
      large areas on the sky, they probably cannot be located very far
      from the Sun and therefore the studies of such narrow-lined clouds
      may give useful information about the gas in the solar
      neighborhood. Usually this gas is studied through corresponding
      absorption lines, which allow estimation of physical conditions in
      the local gas. \ion{H}{i} 21-cm emission line is less useful in
      this respect, but may be still usable for large scale surveys to
      find out possible interesting features in the local neighborhood.

      As a result, we may state that the ring clouds seem to be a rather
      unique feature on the sky. Most likely they are the coldest clouds
      observable in the \ion{H}{i} 21-cm emission line. Also slightly
      warmer clouds (clouds with slightly wider 21-cm emission lines)
      may be related to rather local gas structures inside or near the
      LB. There is some probability that also the gas from $(l =
      320\degr, b = -10\degr)$ to $(l = 260\degr, b = -55\degr)$ may be
      related to the structure, which we called a ring, but in this case
      in larger scales the structure must considerably deviate from a
      perfect planar ring. Some properties of all these clouds may be
      similar to those of the \ion{H}{i} self-absorption features,
      observed predominantly near the galactic plane, where our approach
      to the emission data is most likely not applicable, but in this
      paper we have not studied this in detail to make firm statements.

   \begin{acknowledgements}
      The author would like to thank W.~B.~Burton for providing the
      preliminary data from the LDS for program testing prior the
      publication of the survey. A considerable part of the work on
      creating the decomposition program was done during the stay of
      U.~Haud at the Radioastronomical Institute of Bonn University (now
      Argelander-Institut f\"{u}r Astronomie). The hospitality of the
      staff members of the Institute is greatly appreciated. We thank
      drs. I. Kolka, E. Saar, P. Tenjes and K. Annuk for fruitful
      discussions. We also thank our anonymous referee whose
      suggestions have improved the clarity of the paper. The project
      was supported by the Estonian Science Foundation grant no. 7765.
   \end{acknowledgements}

\end{document}